\documentclass[sn-basic,iicol]{sn-jnl}

\usepackage{array}
\usepackage[caption=false,font=normalsize,labelfont=sf,textfont=sf]{subfig}
\usepackage{stfloats}
\usepackage{url}
\usepackage{verbatim}
\usepackage{makecell}
\usepackage{enumitem}

\jyear{2022}%

\theoremstyle{thmstyleone}%
\theoremstyle{thmstyletwo}%

\theoremstyle{thmstylethree}%

\begin{document}

\title[Article Title]{Who is Gambling? Finding Cryptocurrency Gamblers Using Multi-modal Retrieval Methods}


\author[1]{\fnm{Zhengjie} \sur{Huang}}\email{zj.h@zju.edu.cn}

\author*[1]{\fnm{Zhenguang} \sur{Liu}}\email{liuzhenguang2008@gmail.com}

\author[1]{\fnm{Jianhai} \sur{Chen}}\email{chenjh919@zju.edu.cn}

\author[1]{\fnm{Qinming} \sur{He}}\email{hqm@zju.edu.cn}

\author[2]{\fnm{Shuang} \sur{Wu}}\email{wushuang@outlook.sg}

\author[3]{\fnm{Lei} \sur{Zhu}}\email{leizhu0608@gmail.com}

\author[4]{\fnm{Meng} \sur{Wang}}\email{eric.mengwang@gmail.com}

\affil*[1]{\orgdiv{College of Computer Science and Technology}, \orgname{Zhejiang University}, \orgaddress{\city{Hangzhou}, \country{China}}}

\affil[2]{\orgname{Nanyang Technological University}, \orgaddress{\city{Singapore}, \country{Singapore}}}

\affil[3]{\orgdiv{School of Information Science and Engineering}, \orgname{Shandong Normal Unversity}, \orgaddress{\city{Jinan}, \country{China}}}

\affil[4]{\orgdiv{School of Computer Science and Information Engineering}, \orgname{Hefei University of Technology}, \orgaddress{\city{Hefei}, \country{China}}}

\abstract{With the popularity of cryptocurrencies and the remarkable development of blockchain technology, decentralized applications emerged as a revolutionary force for the Internet. Meanwhile, decentralized applications have also attracted intense attention from the online gambling community, with more and more decentralized gambling platforms created through the help of smart contracts. Compared with conventional gambling platforms, decentralized gambling have \emph{transparent rules} and a \emph{low participation threshold}, attracting a substantial number of gamblers. In order to discover gambling behaviors and identify the contracts and addresses involved in gambling, we propose a tool termed \textit{ETHGamDet}. The tool is able to automatically detect the smart contracts and addresses involved in gambling by scrutinizing the smart contract code and address transaction records. Interestingly, we present a novel LightGBM model with memory components, which possesses the ability to learn from its own misclassifications. As a side contribution, we construct and release a large-scale gambling dataset at https://github.com/AwesomeHuang/Bitcoin-Gambling-Dataset to facilitate future research in this field. Empirically, \textit{ETHGamDet} achieves a F1-score of 0.72 and 0.89 in \emph{address classification} and \emph{contract classification} respectively, and offers novel and interesting insights.}




\maketitle

\section{Introduction}\label{sec1}

\textbf{Blockchain} is essentially a distributed ledger that is shared among the nodes in a peer-to-peer network. The nodes, known as \emph{miners}, record all the transactions that occurred in the network following a consensus protocol. The duplicate ledgers stored in the worldwide nodes ensure that transactions are immutable once recorded, endowing blockchain with \emph{tamper-proof} and \emph{decentralization} nature~\cite{zheng2018blockchain}. Fundamentally, the key feature of blockchain is that it maintains a secure and immutable transaction ledger among the peers that do not trust one another.

A \textbf{smart contract} is simply a program stored on the \emph{blockchain}, which will be executed when its predetermined conditions are satisfied~\cite{szabo1994smart}. Smart contracts encode predefined contract terms into runnable code. The execution of a smart contract cannot be terminated by any participant, so that the terms are strictly followed. Smart contracts make the automatic execution of contract terms possible, ensuring fairness  for all participants. Remarkably, smart contracts have enabled a wide range of applications in various domains, including {distributed exchanges}, {wallets}, {crowdfunding}, and {decentralized gambling}~\cite{guo2016blockchain, norta2016designing, ayed2017conceptual,zichichi2019likestarter,qian2019digital}. 

Conventional gambling applications have long suffered from notorious issues such as {opaque guessing processes},  {fictitious prize pools}, and {refusal to pay the winners}. \emph{Smart contracts for gambling}, in contrast, enforce the gambling funds transferred strictly following the predefined rules, making the entire process completely transparent. This resolves the key concerns in the traditional gambling. Consequently, the number of decentralized gambling has seen a skyrocketing growth in the past years.

Discovering \emph{gambling contracts} and \emph{addresses} amongst millions of entities in the blockchain network not only enables us to perceive the security risks, but also offers a grand picture of the whole ecosystem. However, automatically identifying gambling contracts and addresses is by no means easy, and to the best of our knowledge there is still no such framework available.

\begin{figure}
    \centering
    \includegraphics[width=1.02\linewidth]{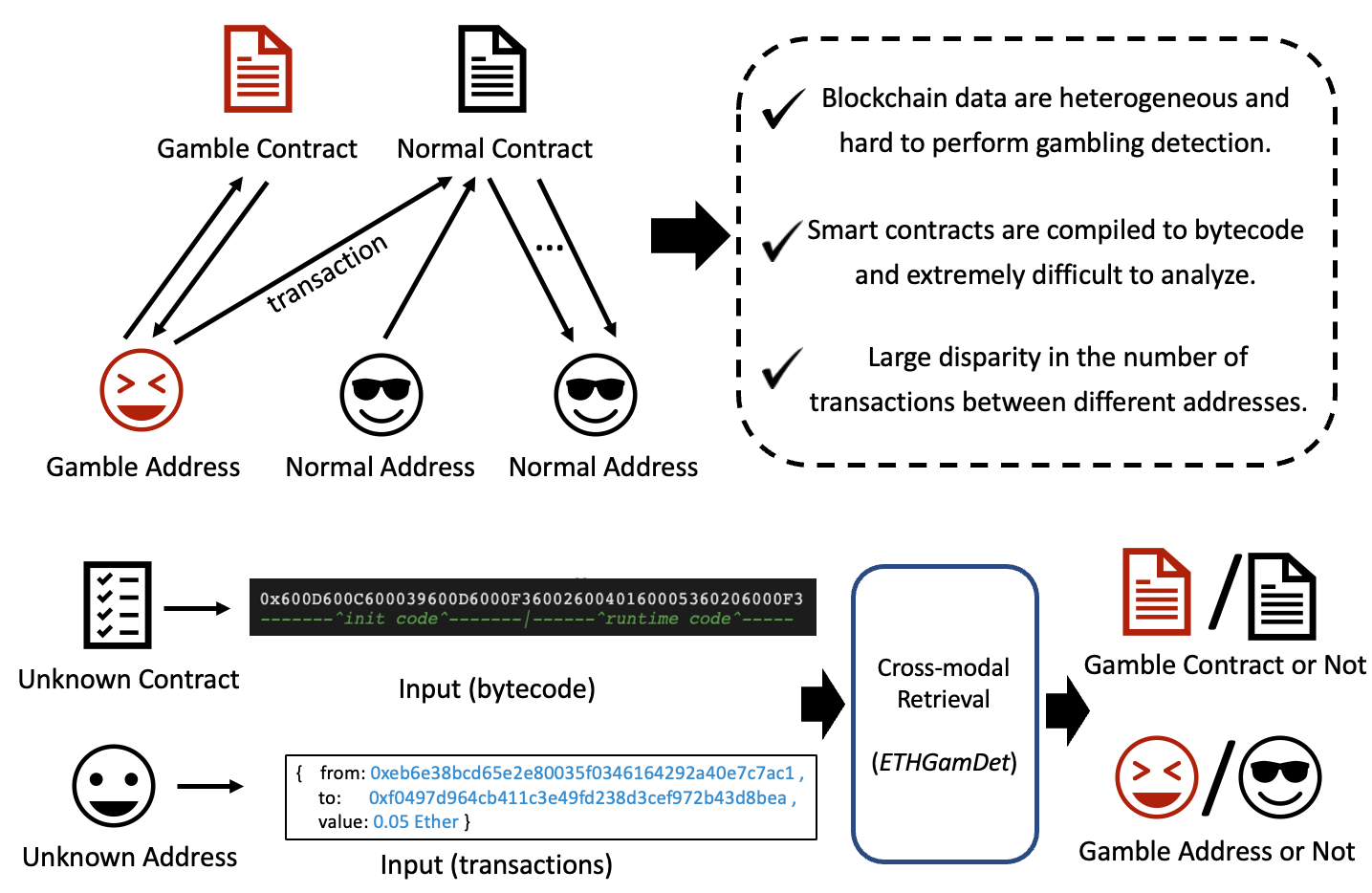}
    \caption{The key challenges of gambling behavior analysis. There are mainly two types of entities in the blockchain platform, addresses and contracts, which are connected by transactions. Our goal is to identify the entities involved in gambling.}
    \label{fig:intro}
\end{figure}

Upon scrutinizing and experimenting with the multi-modal and complex data in Ethereum, we observe that there are three key challenges to be addressed in the gambling detection task, as shown in Fig.~\ref{fig:intro}. \textbf{(1)} The data, such as \emph{transactions, accounts}, and \emph{smart contracts}, in blockchain is heterogeneous and multi-modal, it is still unclear how to formulate the data for gambling-related knowledge extraction. \textbf{(2)} Smart contracts are compiled to bytecode before their deployment on Ethereum. The bytecode, which is in the form of assembly language, is extremely difficult to understand and process. \textbf{(3)} The scales of transaction data of different addresses vary greatly, making it inherently challenging to construct unified and effective features for all addresses.

In this paper, we seek to formulate the gambling detection problem and present the first framework for it. We also contribute a large-scale multi-modal dataset for this task, hoping to inspire future researchers. We try to tackle the challenges in this problem by advocating three key designs: \textbf{(1) Contract Classification.} We utilize the smart contract bytecode to identify gambling-related smart contracts $\{\textbf{S}_i\}_{i=1}^k$. Specifically, we disassemble the bytecode to obtain the contract operation code. We then devise principled features for the operation code and present a novel classifier, consisting of a machine learning model and a memory component, to identify gambling contracts. 
\textbf{(2) Address Classification.} We collect the transaction records of the gambling-related contracts $\{\textbf{S}_i\}_{i=1}^k$. Sifting through these records, we obtain all the addresses $\{\textbf{A}_i\}_{i=1}^r$ that have transactions with $\{\textbf{S}_i\}_{i=1}^k$. To detect the addresses that are truly involved in gambling, we construct a transaction graph for each address $\textbf{A}_i$. The transaction graph models the money transfer activities between $\textbf{A}_i$ and other addresses, revealing the behavior pattern of $\textbf{A}_i$. We then propose a cutting-edge approach for extracting features from the graphs and perform classification with our proposed classifier.
\textbf{(3) Correction.} Finally, we engage in a feedback correction method for improving the overall accuracy of the system. 
Particularly, we leverage the results of the address classification to refine the classification of smart contracts. 

This paper makes the following contributions:
\begin{itemize}[topsep=1.5pt, leftmargin=\dimexpr\labelwidth + 0\labelsep\relax]
\item We propose a new problem of detecting gambling contracts and addresses from multi-modal data, and present the first strong baseline for the problem. We have constructed and released a large-scale benchmark for evaluating gambling detection approaches, and publicized our code to engage the community. 
\item We provide a novel framework \emph{ETHGamDet} to tackle the problem. Within the framework, we present a LightGBM model with memories to boost the classification accuracy, a feature extraction paradigm for smart contract bytecode, and a feature extraction paradigm for Ethereum addresses to realize the identification of addresses participating in gambling. One key highlight of \emph{ETHGamDet} is the ability to directly handle smart contracts through bytecode (without reliance on access to source code), improving the utility of the system.
\item \emph{ETHGamDet} achieves a 0.72 and 0.89 F1-score on gambling contract and address classification, respectively. Through an empirical study, we also gain interesting insights and findings in this problem. We believe our work is an important step towards more intelligent and insightful gambling detection.
\end{itemize}

\section{Multi-modal Retrieval Method}\label{sec2}

\textbf{Problem Formulation.}\quad Formally, presented with sets of smart contracts $\{S_i\}_{i=1}^m$ and account addresses $\{A_i\}_{i=1}^n$, we are interested in detecting all smart contracts and account addresses involved in gambling. We propose to jointly identify the smart contracts and addresses by  scrutinizing two modalities, namely (1) smart contract code, and (2) the transaction records of the account addresses.

\begin{figure*}
    \centering
    \includegraphics[width=0.98\linewidth]{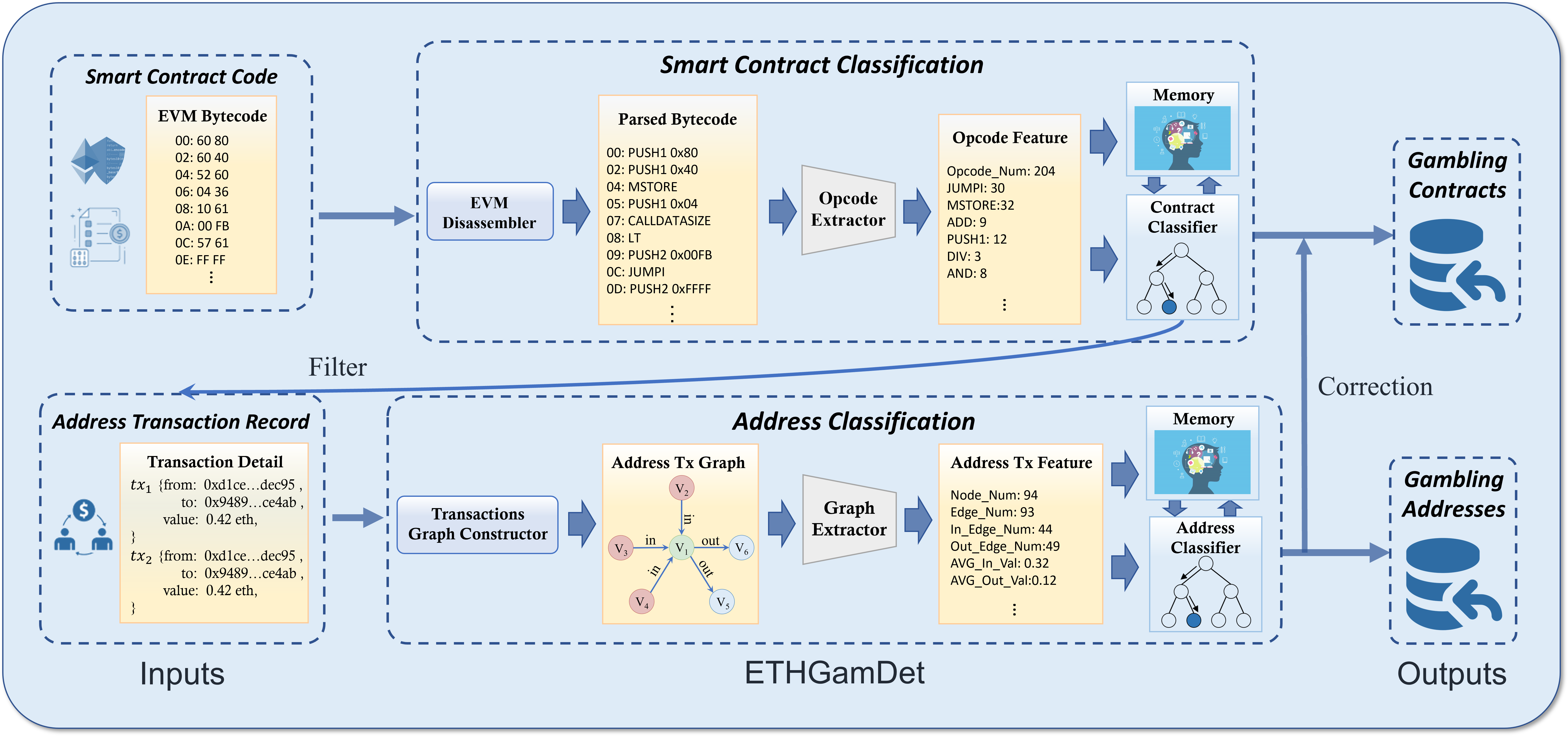}
    \caption{The overview of the \textit{ETHGamDet} system. There are three layers in this systems: \textbf{a) The Inputs Layer} is used to provide input data consisting of Smart Contract Codes and Address Transaction Records to \textit{ETHGamDet}. \textbf{b) The Retrieval Layer} is where \textit{ETHGamDet} processes the input data and outputs the result. This layer uses a multi-modal method to process contract codes and transaction records. This layer is mainly divided into two parts, \textit{Smart Contract Classification} and \textit{Address Classification}. After completing the data processing of these two parts, the system will also use the address classification results to refine the contract classification results, for reducing false positives and improving the accuracy of the system. \textbf{c) The Outputs Layer} receives the classification results of \textit{ETHGamDet}. After being processed by the ETHGamDet, we can get the contract data and address data involved in gambling.}
    \label{fig:Arch}
\end{figure*}

\textbf{Method Overview.}\quad The proposed method \textit{ETHGamDet} (\underline{Eth}ereum \underline{Gam}ble \underline{Det}ector) is outlined in Fig.~\ref{fig:Arch}, consisting of three key components: (1) smart contract classification, (2) account address classification, (3) feedback correction for reducing the false positive rate of smart contract classification. Next, we will introduce the three components, respectively.

\subsection{Smart Contract Classification}
\label{sec:Extraction of Smart Contract Codes}
Presented with millions of smart contracts, we must tackle the following challenges to detect gambling ones. (1) Within the contracts, only a small fraction of them are involved in gambling. (2) Most contracts are not open sourced, which means we have only their bytecode in hand.

Naively, we may directly train a classification model that takes bytecode as input and the label as output. However, the bytecode is a binary code sequence, and it is extremely difficult for a classification model to directly reason and understand the underlying semantics from the bytecode, leading to low accuracy. Another solution is to decompile the bytecode into source code, and adopt off-the-shelf approaches~\cite{grech2019gigahorse, suiche2017porosity, zhou2018erays, albert2018ethir} for contract classification. However, it is well known that reverting the bytecode back to the source code is extremely difficult and will inevitably introduce wrong translations~\cite{liu2019survey}.

Therefore, we propose to disassemble the bytecode into EVM opcodes, and cast EVM opcodes of different lengths into unified features. As demonstrated in Fig.~\ref{fig:Arch}, our smart contract classification component consists of three modules: EVM Disassembler, Opcode Feature Extractor and Contract Classifier.

\subsubsection{EVM Disassembler}

Since most smart contracts exist in the form of bytecode, we implement a EVM disassembler to extract EVM opcodes from them.

Specifically, bytecode is an executable program, consisting of a sequence of $\langle opcode,~data \rangle$ pairs.
The Ethereum Virtual Machine (EVM) is the runtime environment for bytecode on Ethereum. In order to analyze the execution logic of the smart contract, we need to convert bytecode to assembly language using the opcode rules of EVM.

The EVM opcode is divided into specific instruction sets such as arithmetic operations, logical and comparison operations, control flow, system calls, stack operations, and memory operations~\cite{wood2014ethereumyellow}. In addition to typical bytecode operations, the EVM must manage account information (\emph{e.g.,} addresses and balances), current gas prices, and block information~\cite{hildenbrandt2018kevm}. Table~\ref{tab:1} lists several common Ethereum opcodes. Each opcode corresponds to a specific assembly language operation.

\subsubsection{Opcode Feature Extractor}

After disassemble the bytecode of a smart contract, we obtain its corresponding opcode sequence. Since opcode sequence is unstructured data and different opcode sequences of different contracts have distinct lengths, we need to further encode the opcode sequence into a fixed-length feature vector.

We investigated all the 136 opcode operations in EVM one by one, and count the number of occurrences of each opcode in each smart contract. In the experiments, we obtain two empirical insights. (1) Different from other smart contracts, contracts for gambling frequently involve \emph{random number generation} operations and  \emph{gambling fund collection} operations. (2) We further observe that some opcodes appear rarely in contracts (\emph{e.g.,} PUSH5-PUSH32, DUP5-DUP16, SWAP5-SWAP16, etc.). Based on these observations, we discard the opcode operations that are both rarely used and unrelated to gambling. Finally, we settle down to 80 opcode operations, and utilize the number of occurrence of each opcode operation in the contract as the features.

\subsubsection{Contract Classifier}
Following the EVM Disassembler and Opcode Feature Extractor, we map each smart contract to a feature vector. We then engage a classification model to distinguish feature vectors for gambling contracts from that of other contracts. After scrutinizing  the existing methods, we select LightGBM~\cite{ke2017lightgbm} as our contract classifier model. We would like to highlight that in order to make the model more intelligent, we innovatively augment LightGBM with an additional memory component.

Specifically, LightGBM uses the negative gradient of the loss function as the residual approximation of the current decision tree, and then uses the residual approximation to fit the new decision tree~\cite{friedman2001greedy, chen2015xgboost}.

\begin{table*}

\centering
\caption{EVM opcode examples and their corresponding operations.}

\label{tab:1}       

 \resizebox{0.98\textwidth}{!}{
\begin{tabular}{|c|l|l|}

\hline

\textbf{Opcode} & \textbf{Assembly} & \textbf{Description}  \\
\hline

0x00 & stop & End command \\

0x01 & add & Pop two values, add them, and push the result to the stack \\

0x02 & mul & Pop two values, multiply them, and push the result to the stack \\

0x03 & sub & Pop two values, subtract the two and push the result to the stack \\

0x10 & lt & If the first popped value is less than the second, push 1, otherwise push 0 \\

0x11 & gt & If the first popped value is greater than the second, push 1, otherwise push 0 \\

0x14 & eq & Pop two values, if the two values are equal, push 1, otherwise push 0 \\

0x15 & iszero & Pop one value, if the value is 0, push 1 to the stack, otherwise push 0 \\

0x34 & callvalue & Get the transfer amount in the transaction \\

0x35 & calldataload & Get the value of the input field in the transaction \\

0x36 & calldatasize & Get the length of the input field in the transaction \\

0x50 & pop & Pop the top value from the stack \\

0x52 & mstore & Pop two values arg0 and arg1 in turn, and store arg1 at arg0 in memory \\

0x54 & sload & Pop the value as the storage index, and load the value to the stack \\

\hline

\end{tabular}}

\end{table*}

Mathematically, The LightGBM can be viewed as an additive model consisting of $K$ trees:
\begin{equation}
\small
\hat{y}_i = \sum_{k=1}^{K}f_{k} (x_{i}), f_{k}\in F
\end{equation}
The model is optimized using the \emph{Boosting} algorithm. Specifically, it starts with a constant prediction and learns a new function each time, which is given by:
\begin{align}
\small
\hat{y}_i^0 &= 0 \\
\hat{y}_i^1 &= f_{1}(x_i) = \hat{y}_i^0 + f_1(x_i) \\
\hat{y}_i^2 &= f_{1}(x_i) + f_{2}(x_i) = \hat{y}_i^1 + f_2(x_i) \\
&\dots \nonumber\\
\hat{y}_i^t &= \sum_{k=1}^{t}f_{k} (x_{i}) = \hat{y}_i^{t-1} + f_t(x_i)
\end{align}
At step $t$, the objective function can be formulated as:
\begin{align}
\small
Obj^{(t)} &= \sum_{i=1}^{n}l(y_i,\hat{y}_i^t) + \sum_{i=i}^{t}\Omega (f_{i}) \\
&= \sum_{i=1}^{n}l(y_i,\hat{y}_i^{t-1}+f_t(x_i)) + \Omega (f_{t}) + constant \nonumber
\end{align}
The \emph{residual} at step $t-1$ is defined as:
\begin{equation}
\small
res=\hat{y}_i^{t-1} - y_i,
\end{equation}
For each step of LightGBM, we only need to fit the residual of the previous step when generating the decision tree.

Empirically, we found that LightGBM is nearly 10 times faster than XGBoost, requires only one-sixth of the memory used by XGBoost, and more notably, has improved accuracy. LightGBM can speed up the training of the GBDT model without compromising the accuracy due to the following improvements made~\cite{ke2017lightgbm}:
Gradient-based One-Side Sampling (GOSS) and Exclusive Feature Bundling (EFB).

\textbf{Gradient-based One-Side Sampling (GOSS).}\quad is an algorithm that balances reducing the amount of training data and ensuring accuracy. GOSS achieves the purpose of improving efficiency by distinguishing instances with different gradients, retaining instances with larger gradients and randomly sampling instances with smaller gradients.

\textbf{Exclusive Feature Bundling (EFB).}\quad reduces feature dimension by feature bundling. EFB has designed an indicator termed conflict rate, which is used to measure the degree of mutually exclusiveness of features. When this metric is small, we can bundle the two features without affecting the final accuracy since they are not completely mutual exclusive. 

\textbf{Memory Component.}\quad Conventionally, LightGBM is composed of purely a classification model with a set of trained parameters. Inspired by the structure of human brain, our LightGBM system is designed to possess two components, \emph{a classification model} and \emph{a memory}. The memory remembers the misclassified samples. We periodically replay the misclassified samples to update the model parameters so that the classification model constantly learns from its own mistakes. This is consistent with our brain mechanism that occasionally replays important events registered in our memory. In particular, our LightGBM system is formulated in Algorithm~\ref{alg:memory}. Briefly, the process can be sketched as:
\begin{itemize}[topsep=2pt, leftmargin=\dimexpr\labelwidth + 0.5 \labelsep\relax]
\item[(1)] In the first-round training, we train the classification model of our LightGBM  using all the training samples.
\item[(2)] We pick all the wrongly classified samples of the trained classification model and their corresponding labels, putting them into the memory component.
\item[(3)] In the second-round training, we re-train the classification model using all the training samples. However, after every $k$ epochs, we add an additional epoch for replaying wrongly classified samples in the memory. More specifically, we (a) pick all the samples in the memory and (b) randomly select an equal number of samples from the training set to form the \emph{training subset} for the replay epoch.
\item[(4)] Iterate (2) and (3) till converge.
\end{itemize}

Through replaying misclassified data, our LightGBM is endowed with the ability to intelligently learn and remember its own mistakes, ultimately improving the classification accuracy. 

\begin{algorithm}[t]
 \small   
\caption{LightGBM model with memory component}\label{alg:memory}
\textbf{function train():} \\
\textit{Input}: gambling train dataset : \textbf{train\_data}\\
\textit{Output}: predict model : \textbf{model}
\begin{algorithmic}[1] 
\State \textbf{let} memory = [ ]
\State \textbf{let} model = LightGBM.initialize()
\While{pred\_loss $<$ loss\_threshold}
\State input = train\_data + (each $k$ epochs) memory
\State model, pred\_loss, err\_data = model(input)
\State memory = memory + err\_data
\EndWhile
\State \textbf{return} model
\end{algorithmic}

\textbf{function predict():} \\
\textit{Input}: gambling test dataset : \textbf{test\_data}\\
\textit{Input}: predict model : \textbf{model}\\
\textit{Output}: Predict Results : \textbf{results}
\begin{algorithmic}[1] 
\State \textbf{let} results = [ ]
\For{data in test\_data}
\State pred = model.predict(data)
\State results.append(pred)
\EndFor
\State \textbf{return} results
\end{algorithmic}
\end{algorithm}

\subsection{Address Classification}
\label{sec:Extraction of Address Transaction Records}
Up to this point, we are able to classify smart contracts into \emph{gambling contracts} and \emph{others}. For the addresses that have transactions with the contracts classified as gambling ones, we further analyze their behaviors to decide whether they are indeed gamblers. 

Towards this aim, we first construct a transaction graph for each address. Then, we perform graph feature extraction from the transaction graph  and classify the features using a classification model. On the whole, our proposed address classification consists of three modules: Transaction Graph Constructor, Graph Feature Extractor and Address Classifier.

\subsubsection{Transaction Graph Constructor}
Unlike traditional transactions, transactions on Ethereum are not that complicated. We extract the payment address, recipient address, and transaction amount as the features of the transaction. For each address $A_i$, we model all the addresses that have transactions with $A_i$ as the nodes and cast the transactions as edges. Particularly, if $A_i$ is the payment address, an output edge is constructed. Conversely, if $A_i$ is the recipient address, an input edge is added. The weight of the edge is set as the transaction amount.     

Constructing such a transaction graph for each address have the following benefits. (1) Different from other addresses, gambling addresses often have multiple transactions target to the same recipient address, \emph{i.e.,} repeatedly adding funds to the identical gambling pool. This bizarre behavior is often not easily captured by conventional feature vectors but can be easily revealed by this graph structure. (2) The activeness of different addresses varies much, leading to a large variation in the number of transactions for each address. This graph enables us to devise fix-length features for all addresses.

\begin{figure}[ht]
    \centering
    \includegraphics[width=0.8\linewidth]{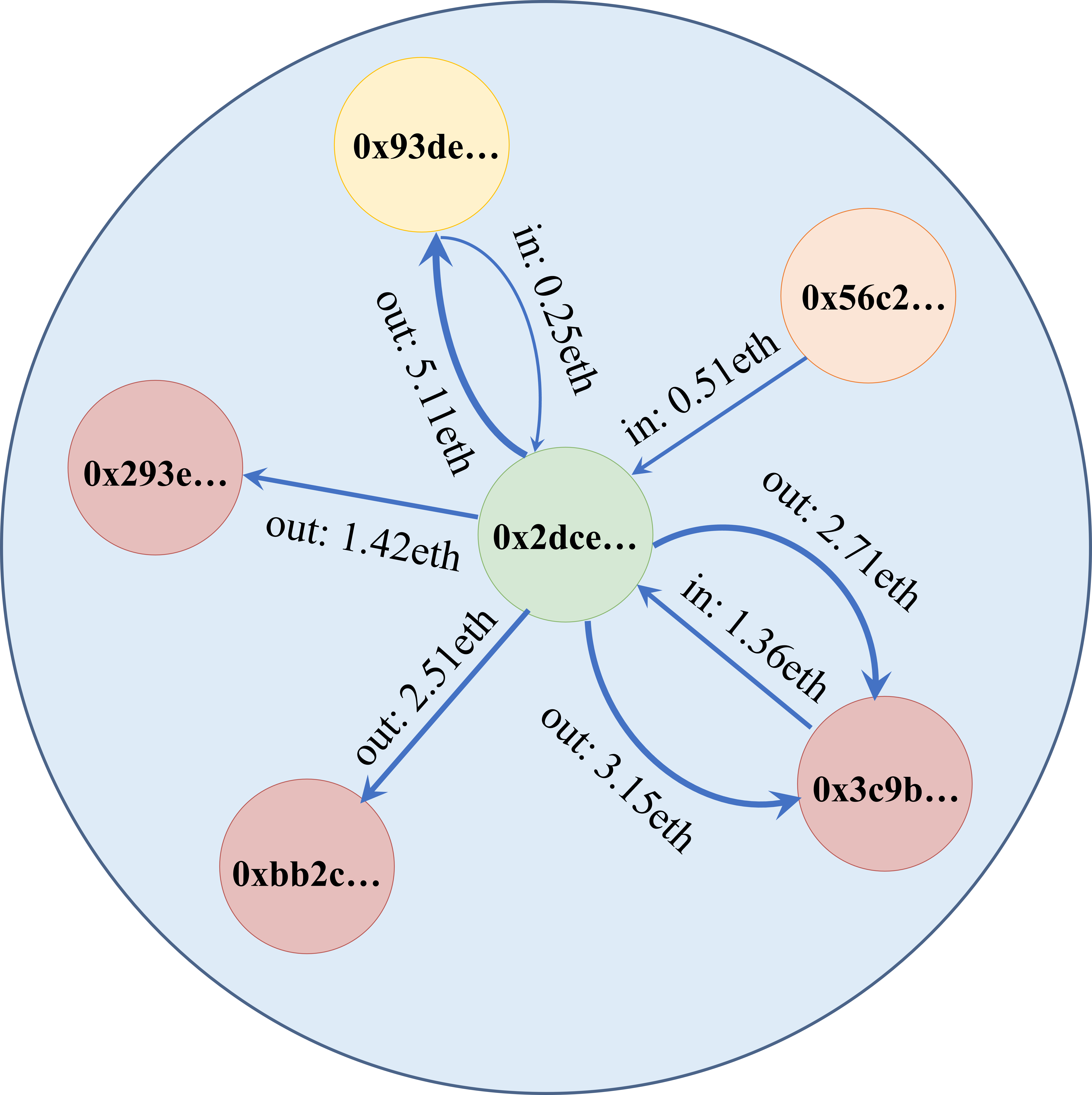}
    \caption{An example of an address transaction graph. Each address has a transaction graph, which records all addresses that have transactions with that address. Nodes in the graph represents addresses, the edges between nodes represent transactions, and the direction of the edges represents the flow of the money.}
    \label{fig:graphexp}
\end{figure}

Fig.~\ref{fig:graphexp} shows an exemplar transaction graph. For the address located in the center of the figure, its transaction graph contains a total of seven edges. Two input edges represent two money incoming transactions  and five output edges stands for five money outgoing transactions. In total, the address have transactions with 5 other addresses. It is worthy mentioning that there might be multiple transactions between two addresses and each transaction is modeled as an edge.

In order to construct and process large-scale graph data, we choose Neo4j as the graph database. Neo4j is a high-performance NoSQL database capable of storing structured data on the graph network~\cite{miller2013graph, webber2012programmatic}.

\begin{table*}[!h]
\small
\caption{Metrics used for graph feature extraction}
\label{tab:2}
\centering
\begin{tabular}{|c|l|l|}
\hline
\textbf{Metrics Type} & \textbf{Metric Name} & \textbf{Description} \\
\hline
    \multirow{4}{*}{Basic Metrics}
    & vertex\_number & The total number of vertices in the entire graph \\
    & edge\_number & The total number of edges in the entire graph \\
    & in\_edge\_number & The total number of input edges in the entire graph \\
    & out\_edge\_number & The total number of output edges in the entire graph \\

\hline
    \multirow{3}{*}{Degree Metrics}
    & vertex\_degree & The average degree of the nodes in the graph \\
    & vertex\_in\_degree & The average in-degree of the nodes in the graph \\
    & vertex\_out\_degree & The average out-degree of the nodes in the graph \\

\hline
    \multirow{9}{*}{Amount Metrics}
    & total\_amount & The sum of the amount of all transactions \\
    & total\_in\_amount & The sum of the amount of all input transactions \\
    & total\_out\_amount & The sum of the amount of all output transactions \\
    & avg\_amount & Average amount of all transactions \\
    & avg\_in\_amount & Average amount of all input transactions \\
    & avg\_out\_amount & Average amount of all output transactions \\
    & amount\_variance & Amount variance for all transactions \\
    & in\_amount\_variance & Amount variance for all input transactions \\
    & out\_amount\_variance & Amount variance for all output transactions \\

\hline
\end{tabular}
\end{table*}

\subsubsection{Graph Feature Extractor}
To extract features from the transaction graph, we engage in three kinds of metrics, namely \emph{basic metrics}, \emph{degree metrics}, and \emph{amount metrics}, which are demonstrated in Table II.

\emph{Basic Metrics} mainly focus on the overall structure of the graph, which capture the number of nodes, edges, and edge types in the transaction graph.

\emph{Degree Metrics} concern the nodes of the graph. The degree metrics mainly model the average degree, average in-degree, and average out-degree of nodes. 

\emph{Amount Metrics} concentrate on the edges of the graph. 
Amount metrics mainly include the sum of transaction amounts, average transaction amount, and transaction amount variance for the edges.

\subsubsection{Address Classifier}
After obtaining the transaction graph features for each address, we use a machine learning model to classify the addresses. For the machine learning model, we adopt the LightGBM model with memory component introduced before in Subsection~\ref{sec:Extraction of Smart Contract Codes}. Put differently, we reuse the structure of the smart contract classifier but train it with  address data.

\subsection{Feedback Correction}
\label{sec:Feedback Correction}
Due to the huge number of contracts on Ethereum, gambling contracts only account for a tiny portion of them. Therefore, even a small false positive rate of the smart contract classifier would lead to a large number of non-gambling addresses classified as gambling ones. Motivated by this, we design a feedback correction mechanism to further reduce the false positive rate of the smart contract classifier.

Empirically, if most of the addresses associated with a contract are not gambling addresses, then the contract is probably not a gambling contract.  After horizontal comparison in experiments, we finally select the threshold 80\%. If less than 80\% of the addresses associated with the contract are gambling addresses, then the contract is likely to be a false positive sample. We revise its label to non-gambling. After adopting this method, the false positive rate of the smart contract classifier is greatly reduced.

\section{Experiments}\label{sec3}
In this section, we evaluate the performance of the model.
We seek to answer the following research questions:
\begin{itemize}[topsep=1.5pt, leftmargin=\dimexpr\labelwidth + 0\labelsep\relax]
\item \textbf{RQ1:} How to build the benchmark dataset for evaluating  gambling classification models?
\item \textbf{RQ2:} Is the proposed method effective in identifying gambling smart contracts and addresses? How does it compare to other classification models.
\item \textbf{RQ3:} How is the importance of different features? Can we gain new insights from the empirical results?
\end{itemize}

Next, we first present the experimental settings, followed by answering the above research questions one by one.

\subsection{Experimental Settings}
~\label{sec:exp1}

\subsubsection{Implementation details}
All experiments are done on a server equipped with an Intel E5-2630 v4 2.20GHz CPU and 128G of memory.
The \textit{ETHGamDet} tool was developed using Python as the programming language. For the transaction graph of addresses, we use Neo4j as the database to store and manage the data.

\subsubsection{Evaluation metrics}
\textit{ETHGamDet} mainly classifies whether a contract is providing gambling service or an address is a gambler, which concerns a binary classification problem. Therefore, we adopt metrics commonly used in binary classification tasks: accuracy, precision, recall, and F1-score. 

\subsection{Datasets (RQ1)}
\label{sec:exp2}
As Fig.~\ref{fig:Arch} shows, we need to collect the smart contract code and address transaction records. We observe that the Ethereum browser website \textit{etherscan.io}~\cite{team2017etherscan} provides relevant APIs to obtain contract data and transaction data, so we obtain data through the website's APIs.

It is worth noting that gambling is only one of the various applications on blockchain. Therefore, gambling contracts and gambling addresses only account for a small portion of contracts and addresses in Ethereum. To make the  classification models  aware of this phenomenon, we constructed an imbalanced dataset, where gambling contracts and addresses only accounts for around 20\%.

For model evaluation and training, we constructed two datasets, gambling contract dataset and gambling address dataset~\cite{bitcoingamblingdataset}, respectively. We manually labeled the addresses in our dataset. In total, four annotators participated in the data collection and labeling. We explored a large number of decentralized gambling sites and recorded the contracts published on the sites. By analyzing these contracts and associated transactions, we were able to obtain the contracts and addresses involved in gambling. Table~\ref{tab:3} shows the statistics of the datasets. Interestingly, we would like to point out we also collected 2,585 unlabeled contracts and 28,919 unlabeled addresses in the datasets, leaving a room for exploring semi-supervised classification models on this open problem.

\begin{table}
\centering
\caption{The quantitative information of dataset.}
\label{tab:3}
 \resizebox{0.48\textwidth}{!}{
\begin{tabular}{ccccc}
\hline\noalign{\smallskip}
\textbf{Dataset} & \textbf{gamble} & \textbf{non-gamble} & \textbf{unlabel}  & \textbf{total}  \\
\noalign{\smallskip}\hline\noalign{\smallskip}
Contract & 260 & 1,040 & 2,585 & 3,885 \\
\noalign{\smallskip}
Address & 10,423 & 51,004 & 28,919 & 90,346 \\
\noalign{\smallskip}\hline
\end{tabular}}
\end{table}

\subsubsection{Gambling Contract Dataset}
We collected a total of 260 gambling smart contracts from  decentralized gambling websites, such as Dicether~\cite{dicether}, Degens~\cite{degens}. At the same time, in order to construct the negative samples required for training, we selected 1,040 smart contracts that are not involved in gambling (\emph{e.g.}, erc20~\cite{victor2019measuring}, erc721~\cite{chirtoaca2020framework}, mixer~\cite{feng2019survey}, etc.). In the dataset, we use accounts to refer to contracts (e.g. 0x3fe2b...f8a33f), where 1, 0, and -1 to represent the gamble, non-gamble, and other types, respectively.

\subsubsection{Gambling Address Dataset}
We collected 10,423 gambling addresses that have transactions with gambling contracts. Moreover, we also selected 51,004 non-gambling addresses (such as exchanges~\cite{warren20170x}, wallet addresses~\cite{er2017blockchain}, etc.), making the gambling address dataset more complete. In the dataset, we use accounts to refer to addresses (e.g. 0xd1ce...edec95), where 1, 0, and -1 represent the gamble, non-gamble, and other types, respectively.

\begin{table*}[!h]
\renewcommand\arraystretch{1.1}
\scriptsize
\caption{\textit{ETHGamDet} Experimental Results}
\label{tab:4}
\centering
 {
\begin{tabular}{|c|c|c|c|c|c|}
\hline
\textbf{Classification Task} & \textbf{Model Selection} & \textbf{Accuracy} & \textbf{Precision} & \textbf{Recall} & \textbf{F1-score} \\
\hline
    \multirow{9}{*}{Address Classification}
    & LR & 0.53 & 0.16 & 0.41 & 0.23 \\
    & SVM & 0.65 & 0.25 & 0.52 & 0.34 \\
    & KNN & 0.64 & 0.24 & 0.53 & 0.33 \\
    & Bernoulli NB & 0.55 & 0.17 & 0.43 & 0.24 \\
    & Gaussian NB & 0.55 & 0.17 & 0.44 & 0.25 \\
    & Decision Tree & 0.83 & 0.49 & 0.56 & 0.52 \\
    & Random Forest & 0.84 & 0.54 & 0.56 & 0.55 \\
    & LightGBM & 0.88 & 0.63 & 0.76 & 0.69 \\
    & \textbf{ETHGamDet} & \textbf{0.90} & \textbf{0.67} & \textbf{0.77} & \textbf{0.72} \\

\hline
    \multirow{9}{*}{\makecell[c]{Contract Classification \\ (without Correction)}}
    & LR & 0.64 & 0.30 & 0.57 & 0.39 \\
    & SVM & 0.74 & 0.41 & 0.65 & 0.50 \\
    & KNN & 0.78 & 0.47 & 0.78 & 0.59 \\
    & Bernoulli NB & 0.65 & 0.31 & 0.59 & 0.41 \\
    & Gaussian NB & 0.66 & 0.31 & 0.60 & 0.41 \\
    & Decision Tree & 0.81 & 0.52 & 0.85 & 0.65 \\
    & Random Forest & 0.83 & 0.55 & 0.88 & 0.68 \\
    & LightGBM & 0.91 & 0.72 & 0.90 & 0.80 \\
    & \textbf{ETHGamDet} & \textbf{0.92} & \textbf{0.74} & \textbf{0.91} & \textbf{0.82} \\

\hline
    \multirow{9}{*}{\makecell[c]{Contract Classification \\ (with Correction)}}
    & LR & 0.69 & 0.34 & 0.57 & 0.43 \\
    & SVM & 0.81 & 0.52 & 0.65 & 0.58 \\
    & KNN & 0.83 & 0.55 & 0.78 & 0.65 \\
    & Bernoulli NB & 0.70 & 0.35 & 0.59 & 0.44 \\
    & Gaussian NB & 0.70 & 0.35 & 0.60 & 0.44 \\
    & Decision Tree & 0.86 & 0.61 & 0.85 & 0.71 \\
    & Random Forest & 0.88 & 0.65 & 0.88 & 0.75 \\
    & LightGBM & 0.94 & 0.82 & 0.90 & 0.86 \\
    & \textbf{ETHGamDet} & \textbf{0.96} & \textbf{0.88} & \textbf{0.91} & \textbf{0.89} \\

\hline
\end{tabular}}
\end{table*}

\subsection{Classification Results (RQ2)}\label{sec:exp3}

\begin{figure*}[ht]
    \centering
    \includegraphics[width=0.99\linewidth]{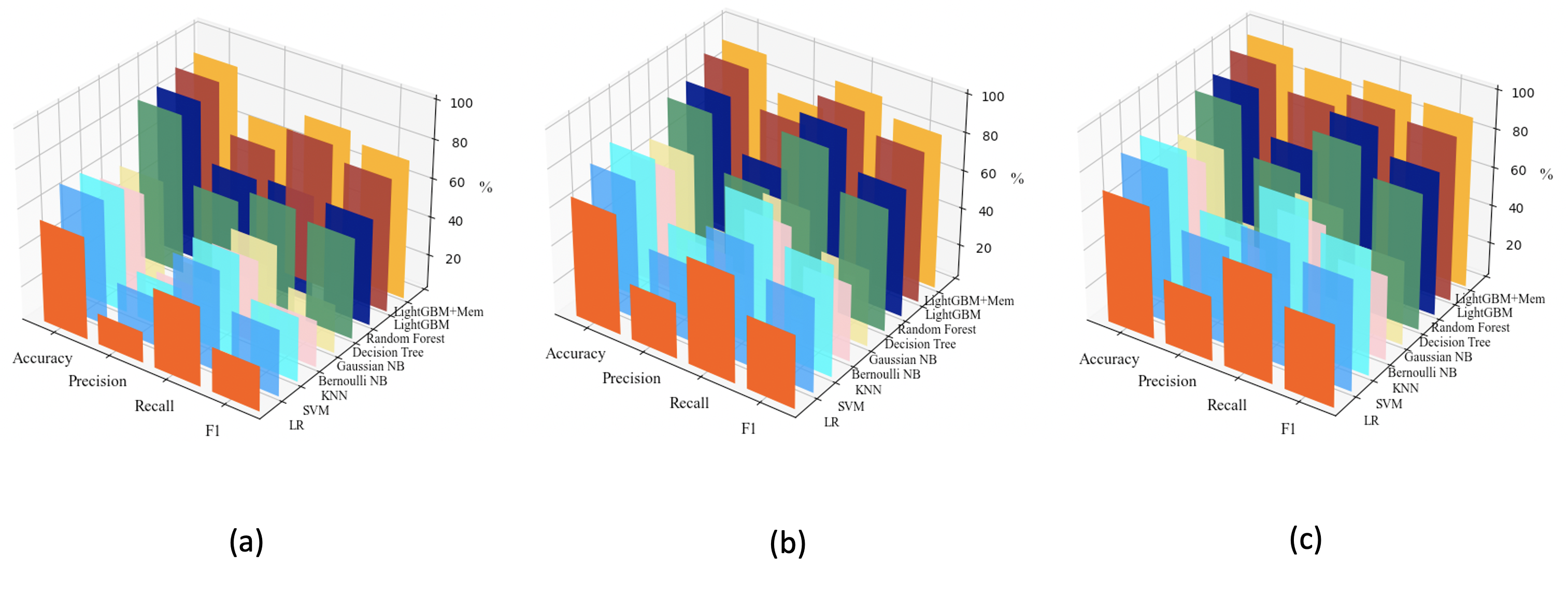}
    \caption{Visually comparison: (a) shows the experimental comparison results of address classification on the gambling address dataset. (b) \& (c) shows the experimental comparison results of contract classification on the gambling contract dataset, while (b) has no feedback correction and (c) has feedback correction. In each graph, the 9 rows from front to back denote the LR, SVM, KNN, Bernoulli NB, Gaussian NB, Decision Tree, Random Forest, LightGBM and LightGBM with memory methods, respectively. For each column in the figures, accuracy, recall, precision, and F1 score are respectively demonstrated from left to right.}
    \label{fig:Visual}
\end{figure*}

In this section, we evaluate the proposed \textit{ETHGamDet} by conducting comparative experiments. Our experiments start from two perspectives: address classification results and contract classification results. By combining these two types of classification results, we have an intuitive understanding of the effect of \textit{ETHGamDet}.

We selected seven commonly used models, LR, SVM, KNN, Bernoulli NB, Gaussian NB, Decision Tree, and Random Forest, for comparison. These models cover the current mainstream machine learning methods. As shown in Table~\ref{tab:4}, experiments demonstrate that the model LightGBM with memory component achieves remarkable results in both classification tasks. In the experiment of identifying gambling addresses, the F1-score of \textit{ETHGamDet} reached 0.72, while in the experiment of identifying gambling contracts, the F1-score of \textit{ETHGamDet} reached 0.89. This suggests that \textit{ETHGamDet} is able to accurately identify the addresses and contracts involved in gambling.

To allow for more intuitive understanding, we also visualized the comparison results in Fig.~\ref{fig:Visual}. The detailed discussions on the experimental results are given below.

\subsubsection{Address Classification Results}
We conducted comparative experiments on a total of 61,427 addresses. The experimental results show that our model consistently outperforms other classification models, and achieved 0.90, 0.67, 0.77, 0.72 in accuracy, precision, recall, and F1-score, respectively. The F1-score of our model is 0.03 higher than the LightGBM without memory component, which ranks the second. 

Interestingly, we observe that all methods tend to have a relatively low precision in the experiments.  We speculate the reasons are two-folds. On one hand, by analyzing the experimental data, we found that there are a large number of addresses having only a few transactions. The behavior of such addresses is extremely difficult to analyze due to the lack of data, affecting the precision of address classification. On the other hand, gambling transactions are similar to legal frequent transfer transactions, which makes it difficult for the models to distinguish them.

\subsubsection{Contract Classification Results}
We conducted comparative experiments on a total of 1,300 labeled contracts. We first compared across different classification models. It can be seen from the experimental results in Table~\ref{tab:4} that our LightGBM algorithm surpasses LR, SVM, Naive Bayes, Decision Tree and Random Forest in various indicators. The empirical evidences reveal that our model is effective in the gambling contract detection task.

Next, we studied the effect of our designed \textit{Correction Feedback} on the performance. 
By default, we used the \textit{Correction Feedback} component to reduce false positives in contract classification. To study its effect, we removed it from the method. After removing \textit{Correction Feedback}, the accuracy, precision, and F1-score of the model all decreased. More specifically, the precision of contract classification decreased from  0.96 to 0.92, and the F1-score of the classification decreased from  0.89 to 0.82.  This reveals the effectiveness of the proposed \textit{Correction Feedback} component.  It is worth noting that since the correction feedback mechanism does not affect the false negative rate of the model, the recall does not change with or without correction feedback.

Finally, we horizontally compare the results of contract classification and address classification. From Table~\ref{tab:4}, we observe that the F1-score of contract classification is generally better than that of address classification. This is because the code in the contract is more semantically rich, and a substantial number of addresses have only a  few transactions. This also shows the rationality of the design of \textit{ETHGamDet}. \textit{ETHGamDet} processes the contract code first and then handles the address transactions. Through this processing pipeline, the better classification results of the contract code can be utilized to the greatest extent.

\subsection{Feature Analysis (RQ3)}
To understand the interpretability of \textit{ETHGamDet}, we perform feature importance analysis in the address classification and the contract classification separately.
We selected the contract classification and address classification models trained in Section~\ref{sec:exp3}, and used the built-in feature importance function of LightGBM to print out the weights of each input feature and draw them into a graph.

\begin{figure}
    \centering
    \includegraphics[width=1\linewidth]{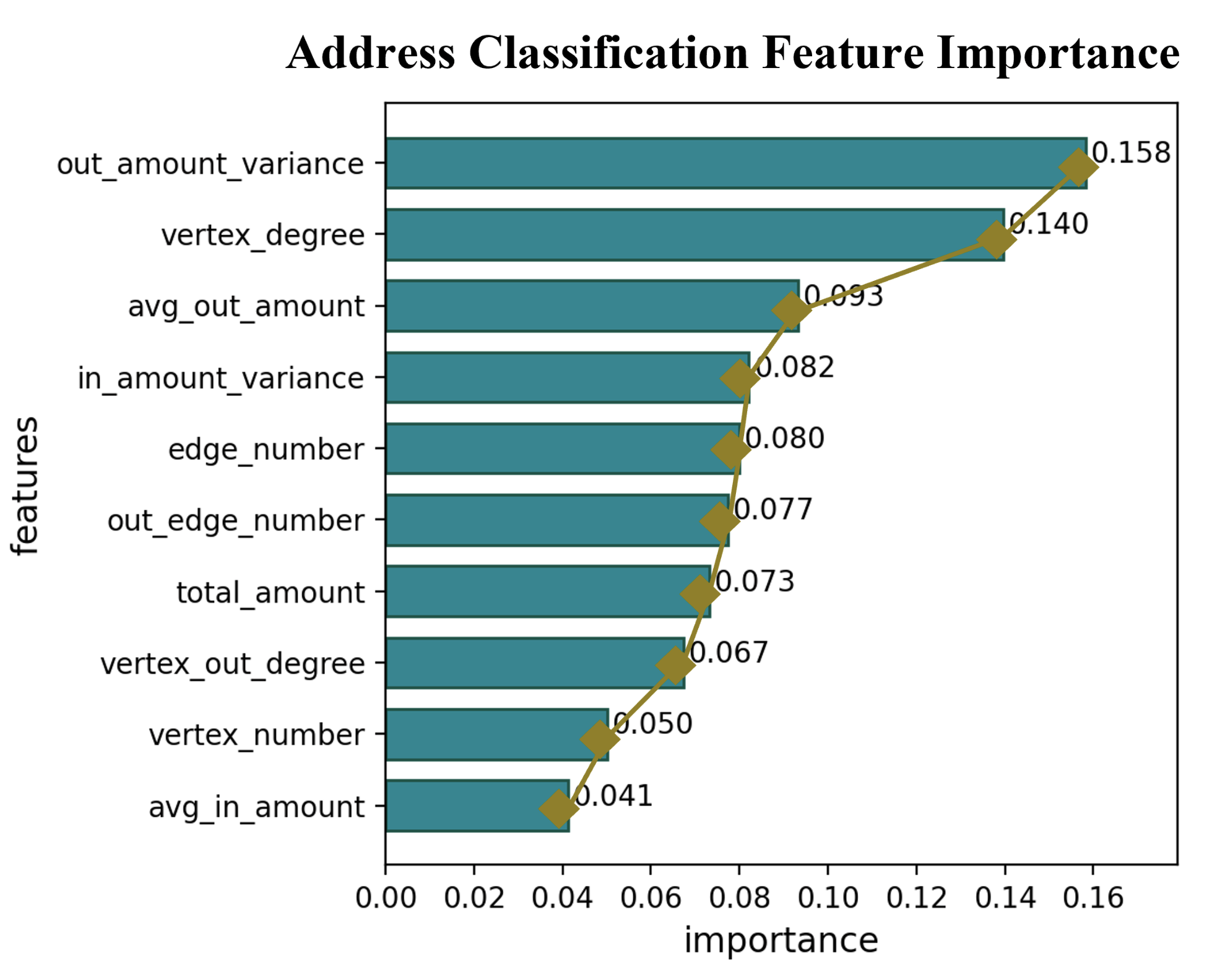}
    \caption{This figure shows the feature importance of address classification. We show the top ten most important features. It can be seen that features \textit{out\_amount\_variance} and \textit{vertex\_degree} are ranked the highest in importance.}
    \label{fig:addressfeature}
\end{figure}

\begin{figure}
    \centering
    \includegraphics[width=1\linewidth]{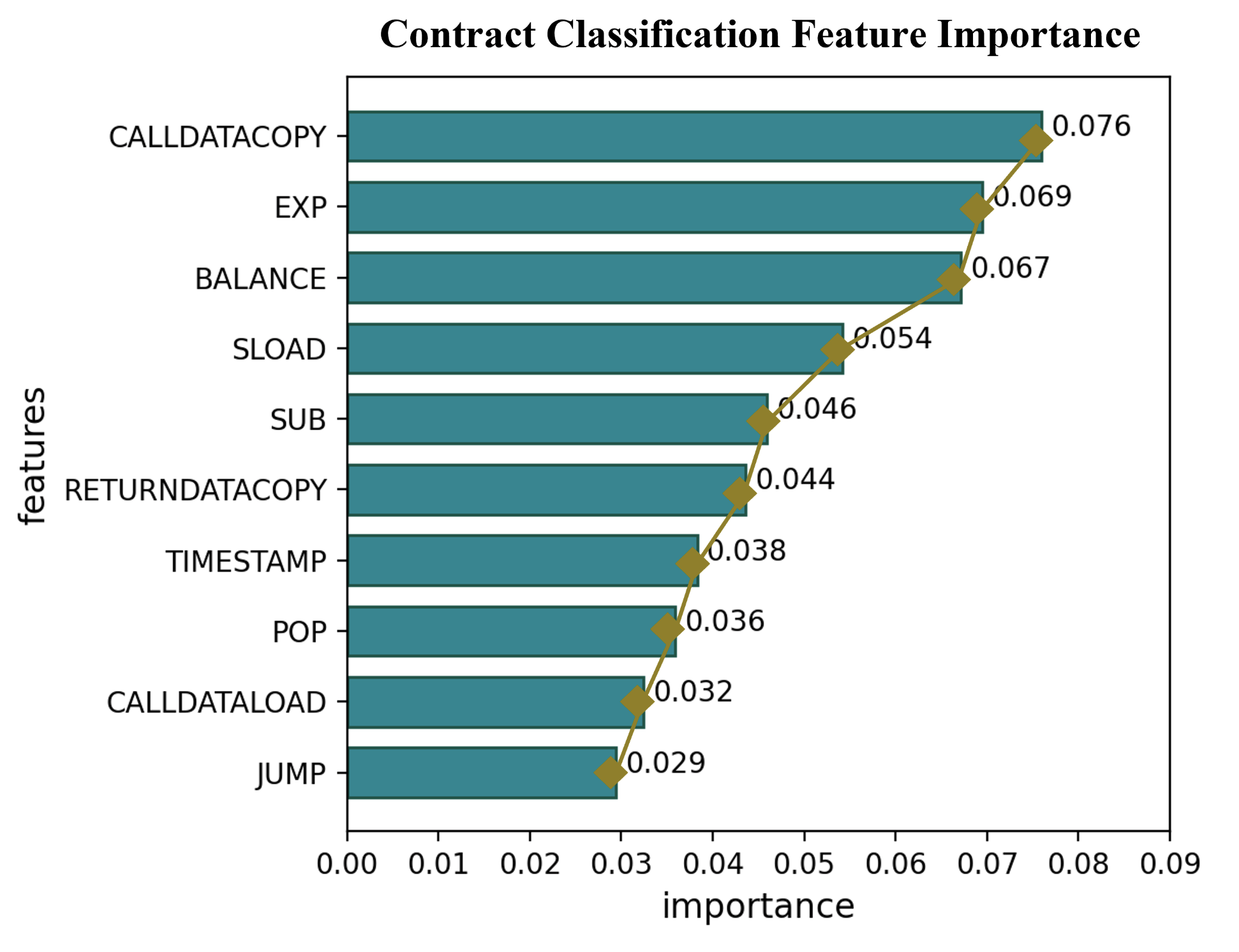}
    \caption{This figure shows the feature importance of contract classification. We show the top ten most important features. It can be seen that features \textit{CALLDATACOPY}, \textit{EXP}, and \textit{BALANCE} are ranked the highest in importance.}
    \label{fig:contractfeature}
\end{figure}

Fig.~\ref{fig:addressfeature} shows feature importance for address classification. From the figure, we have the following observations. (1) Features  \textit{out\_amount\_variance} has the highest weight. The feature \textit{out\_amount\_variance} represents the variance of the amount paid by the gambler to the gambling contract. Unlike behaviors such as transfers, gamblers usually pay a similar amount each time, so gambling addresses are similar in this feature. (2) Another feature with high importance is \textit{vertex\_degree}. This feature can reflect the frequency of gamblers' participation in a particular gamble. We found that gamblers often participate in a certain gambling contract, which is reflected in the graph that there are multiple edges between two nodes.

Fig.~\ref{fig:contractfeature} shows feature importance for contract classification. As can be seen from the figure, CALLDATACOPY, EXP and BALANCE have the highest weights. CALLDATACOPY copies multiple bytes of transaction data into memory. Gambling contracts need to complete bets and withdrawals based on transaction data, so CALLDATACOPY often appears in gambling contracts. Gambling contracts need to generate random numbers based on some data, so the exponentiation modulo operation EXP is often used to generate random numbers. The role of BALANCE is to query the balance in the contract account. The gambling contract needs to judge the balance to calculate the amount of the gambling lottery.

\section{Related Work}\label{sec4}
With the continuous development of blockchain technology, the application ecology based on blockchain is constantly changing the Internet. While these applications gradually expand the use boundaries of the blockchain, they also bring many problems, such as gambling~\cite{scholten2020inside}, money laundering~\cite{campbell2018bitcoin}, darknet transactions~\cite{broadhurst2018malware}, and hacker attacks~\cite{mehar2019understanding}. In recent years, more and more researchers paid attention to these problems and strived to solve them using novel techniques.

Unlike traditional problems, all contract code and transaction records are deployed on the blockchain, which makes it easier for us to obtain all data for analysis. However, the data on the blockchain is multi-modal and lacks  connections to real-world entities. 

At present, most blockchain research is based on two types of data, namely the code of smart contracts and the transaction records of addresses.

By analyzing the code of the smart contract, we may have a better understanding of the purpose or security issues of the contract~\cite{atzei2017survey, chen2017under, fu2019evmfuzzer, mohanta2018overview, macrinici2018smart, grech2018madmax, bhargavan2016formal, brent2018vandal, zhuang2020smart, liu2021combining, liu2021smart}. Chen et al.~\cite{chen2018detecting} used machine learning methods to analyze the contract code, identifying more than 400 contracts that may be related to the Ponzi scheme.
Moreover, analyzing the contract code can also help us find out the loopholes in the contract, so as to avoid being exploited by hackers. Luu et al.~\cite{luu2016making} designed a tool called Oyente. The tool can use the method of symbolic execution to find out four types of vulnerabilities: Transaction-ordering Dependence, Timestamp Dependence, Mishandled Exceptions, and Reentrancy Vulnerability. Kalra et al.~\cite{kalra2018zeus} designed a tool called Zeus by borrowing ideas from the Oyente tool. The tool utilizes both abstract interpretation and symbolic model checking. A more advanced detection tool Securify was proposed by Tsankov et al.~\cite{tsankov2018securify}. The tool first extracts semantic information accurately by analyzing the dependency graph of the contract, and then captures compliance and violation patterns to analyze the security of the contract.

By analyzing the transaction records of the address, we can realize the behavior pattern of the address. Wu et al.~\cite{wu2020phishers} proposed a method to detect phishing scams on Ethereum. This method designs a new network embedding algorithm \textit{trans2vec} to perform data mining on transaction records. Through this method, Wu et al. realized the classification of  addresses in Ethereum, so as to determine whether an address is a phishing one. Chen et al.~\cite{chen2019market} analyzed the leaked transaction records of the Mt.Gox exchange. They model transaction information as a graph, so as to  dig out the abnormal transaction behaviors. Chen et al. found that there was serious market manipulation in the Mt.Gox exchange, illustrating that the regulation of cryptocurrency exchanges needs to be strengthened. 
In addition, address transaction records are also used for abnormal behavior monitoring~\cite{ante2021impact, akcora2020bitcoinheist, lee2019toward, morishima2021scalable, li2021efficient, yan2022blockchain}, such as money flow detection and identity tracing.

Overall, analyzing the smart contract code enables us to comprehend the semantic characteristics of the contract, while analyzing the address transaction records allows us to understand the behavioral characteristics of the address. By analyzing both data sources simultaneously, we are able to capture the vast majority of events on the blockchain. As a result, we resort to a multi-modal retrieval method to process these two types of data to identify the contracts and accounts involved in gambling on Ethereum.

\section{Conclusion and Future Work}\label{sec5}
In this paper, we propose \textit{ETHGamDet}, a novel Ethereum gambling detection system. The tool employs a multi-modal approach to analyze real data on Ethereum. \textit{ETHGamDet} attempts to address the difficulties of gambling detection from three perspectives: (1) \emph{Problem.}~ Divides the issue into two subtasks: classification of gambling contracts and classification of gambling addresses. (2) \emph{Dataset.}~  Constructs smart contract and address transaction datasets so that gambling detection algorithms can be evaluated. (2) \emph{Method.}~  Presents a unique framework for jointly detecting gambling contracts and gambling addresses. Within the framework, it develops paradigms for smart contract feature extraction and address feature extraction, and introduces a new LightGBM model enhanced with memories. Extensive experiments are conducted to assess the proposed method, which demonstrates its efficacy and yields novel findings.

Although \textit{ETHGamDet} has shown exceptional success in identifying gambling-related activities, there are still limitations. In future work, we will enhance \textit{ETHGamDet} in terms of performance and application breadth. (1) \emph{Performance.}~ On the one hand, we will improve the opcode extractor in the smart contract classification phase. We will attempt to conduct feature understanding on smart contracts using program analysis techniques such as control flow graphs, abstract syntax trees. On the other hand, at the phase of address classification, we will enhance the graph extractor. We will try to utilize a graph neural network to perform feature extraction on the constructed graph, in order to mine the features of the transaction graph at a deeper level. (2) \emph{Scope of application.}~ We believe that \textit{ETHGamDet}, as the most advanced gambling identification tool, should not only assist the identification of gambling contracts and gambling addresses, but also distinguish between different types of contracts and addresses. Concerning the gambling contract, we will examine the code of the contract in further detail to evaluate if it is a "fair contract,". Regarding the gambling address, we will study the gambling behavior of the address to establish its gambling preference that whether it is risk-averse or greedy.

With \textit{ETHGamDet}, it is possible to detect and investigate gambling behaviors on Ethereum. By doing further research on these gaming addresses and contracts, we were able to determine the design rules of the gambling game and the preferences of players. We believe our work is a significant advancement in the field of cryptocurrency gambling detection, and we hope it will could inspire the community.

\bmhead{Conflicts of interest}
All the authors have checked the manuscript and have agreed to the submission in International Journal of Multimedia Information Retrieval. There is no conflict of interest.

\bmhead{Data availability}
The datasets analysed during the current study are available in the GitHub repository~\cite{bitcoingamblingdataset}, https://github.com/AwesomeHuang/Bitcoin-Gambling-Dataset.

\bmhead{Acknowledgements}
This work was supported in part by the National Key R\&D Program of China (2021YFB2700500); in part by the National Natural Science Foundation of China (No. 61902348); and in part by the Key R\&D Program of Zhejiang Province (No. 2021C01104).


\bibliography{sn-article}


\end{document}